\documentclass[11pt]{article}
\usepackage[]{inputenc}
\usepackage[T1]{fontenc}
\usepackage{hyperref}
\usepackage{color}
\usepackage{fullpage}

\usepackage{amsmath}
\usepackage{amssymb}

\newcommand{\GG}{\ensuremath{\mathcal{G}}}
\newcommand{\LL}{\ensuremath{\mathcal{L}}}
\newcommand{\HH}{\ensuremath{\mathcal{H}}}
\newcommand{\FF}{\ensuremath{\mathcal{F}}}
\title{On the hyperbolicity of the most general Horndeski theory}
\author{Giuseppe Papallo\footnote{\href{mailto:g.papallo@damtp.cam.ac.uk}{\nolinkurl{g.papallo@damtp.cam.ac.uk}}} \\
\\
\it DAMTP, Centre for Mathematical Sciences, University of Cambridge, \\
\it Wilberforce Road, Cambridge CB3 0WA, UK}

\begin{document}
	\maketitle
	\begin{abstract}
		In this paper we study the hyperbolicity of the equations of motion for the most general Horndeski theory of gravity in a generic ``weak field'' background. We first show that a special case of this theory, namely Einstein--dilaton--Gauss-Bonnet gravity, fails to be strongly hyperbolic in any generalised harmonic gauge. We then complete the proof that the most general Horndeski theory which, for weak fields, is strongly hyperbolic in a generalised harmonic gauge is simply a ``k-essence'' theory coupled to Einstein gravity and that adding any more general Horndeski term will result in a weakly, but not strongly, hyperbolic theory.
	\end{abstract}
	\section{Introduction} % (fold)
		\label{sec:introduction}
		Horndeski theories are the most general four-dimensional diffeomorphism-covariant theories of gravity coupled to a scalar field, with second order equations of motion \cite{Horndeski1974}. Many theories of interest in modern cosmology and astrophysics can be recovered as special cases of Horndeski theories (\emph{k-essence, Galileons, Brans--Dicke, Einstein--dilaton--Gauss-Bonnet, etc.}).

		For these theories to be considered viable alternatives to Einstein's General Relativity (GR), it is necessary that they satisfy some essential mathematical consistency requirements. In particular, as for any other classical theory, it is necessary that the initial value problem be (at least \emph{locally}) well-posed; that is, it should be possible to establish the existence, at least for a small interval of time, of a unique continuous map between the initial data and a solution to the equations of motion.

		Establishing the (local) well-posedness (or lack thereof) of the full non-linear problem is a difficult task. We can somewhat simplify it by considering the equations of motion for non-linear perturbations of a given background solution \(u_{0}\) and noting that for this to be well-posed, it is necessary that all initial value problems obtained by linearising the non-linear problem around any background solution in a neighbourhood of \(u_{0}\) be well-posed. The well-posedness of such linearised problems is closely related to the \emph{hyperbolicity} of the equations of motion. We distinguish between two notions of hyperbolicity, \emph{weak} and \emph{strong}, with the latter implying the former. Roughly speaking, \emph{weak hyperbolicity} means that there are no solutions which grow exponentially fast in time and frequency, while \emph{strong hyperbolicity} corresponds to the existence of an ``energy estimate'' bounding the energy of a solution at a time \(t\) in terms of the energy of the initial data. The existence of such ``energy estimate'' is sufficient to establish, via a standard technique, the local well-posedness of the linearised problem. Moreover, energy estimates constitute a standard tool used in establishing the well-posedness of the non-linear problem.

		The fact that Horndeski theories are diffeomorphism-covariant implies that one will not be able to determine their hyperbolicity unless the gauge is appropriately fixed. The Einstein equations, for example, are strongly hyperbolic in any \emph{generalised harmonic gauge}, while they could fail to be so in other gauges (e.g. they are only weakly hyperbolic in the standard ADM formulation \cite{Arnowitt1959,Nagy2004}).
		
		In addition, while the Einstein equations in generalised harmonic gauge are strongly hyperbolic for \emph{any} choice of background, this is not necessarily true for Horndeski theories. It is in fact known that when the background fields become large, the Horndeski equations may fail to be even weakly hyperbolic (e.g. some cosmological solutions are known to suffer from such pathology \cite{Papallo2017}). 
		On the other hand, it was shown in \cite{Papallo2017} that all Horndeski
		theories are weakly hyperbolic in any generalised harmonic gauge when the background fields are sufficiently ``weak'', i.e. when the equations of motion can be considered as a ``small perturbation'' of the Einstein--scalar-field equations of motion.\footnote{Note that this statement is not trivial, for this perturbation alters the equations of motion at the level of the highest order derivatives and hence could, in principle, alter their character.}

		We will therefore investigate the strong hyperbolicity of the generalised harmonic gauge Horndeski equations linearised around a generic ``weak field'' background. In Ref.~\cite{Papallo2017} it was shown that there exists a \emph{unique} choice of generalised harmonic gauge which makes a subclass of Horndeski theories -- namely, Einstein gravity coupled to ``k-essence'' -- strongly hyperbolic.\footnote{In fact, this theory can be written in \emph{symmetric} hyperbolic form \cite{Rendall2006}.} This theory corresponds to the choice of Horndeski coupling functions \(\GG_{3}=\partial_{X}\GG_{4}=\GG_{5}=0\) (cf. Section~\ref{sec:background} below).

		Furthermore, it was also shown that for another, more general class (\(\GG_{3}\neq 0\), \(\partial_{X}\GG_{4}\neq 0\), \(\GG_{5}=0\)) there exists no choice of generalised harmonic gauge which makes the equations strongly hyperbolic. It was then argued (see Section~IV.D.4 of \cite{Papallo2017}) that any more general Horndeski theory (\(\GG_{5}\neq0\)) would either degenerate into one of the theories already considered (\(\partial_X\GG_{5}=0\)) or would not be strongly hyperbolic in any generalised harmonic gauge (\(\partial_X\GG_{5}\neq0\)).
		
		In this paper we will complete the analysis carried out in \cite{Papallo2017} by providing a rigorous proof of this last claim.
		We will first show that Einstein--Dilaton--Gauss-Bonnet theory, a special case of the general Horndeski theory, fails to be strongly hyperbolic. This, besides being a theory of particular interest on its own, will provide a ``pedagogical'' introduction to the proof of the general result. We will then prove that, in fact, any Horndeski theory including a \(\GG_{5}\neq 0\) term in its action will fail to be strongly hyperbolic around a ``weak field'' background in any generalised harmonic gauge. Hence we will conclude that, indeed, the most general Horndeski theory 
		which is strongly hyperbolic in this setting is simply Einstein gravity coupled to ``k-essence''.
		\medskip

		The paper is organised as follows: in Section~\ref{sec:background} we quickly review the necessary background concepts as well as the results obtained in our previous paper; in Section~\ref{sec:einstein_dilaton_gauss_bonnet_gravity} we prove that Einstein--dilaton--Gauss-Bonnet gravity fails to be strongly hyperbolic in a generic weak field background; in Section~\ref{sec:failure_of_strong_hyperbolicity_for_a_general_horndeski_theory} we show how strong hyperbolicity fails in a general Horndeski theory.
		% section introduction (end)
		% \pagebreak
	\section{Background} % (fold)
		\label{sec:background}
		In this section we will briefly review the necessary background and the results obtained previously. We will refer the reader to \cite{Papallo2017,Sarbach2012} (and references therein) for more exhaustive discussions on these topics. 
		\subsection{Equations of motion} % (fold)
			\label{sub:equations_of_motion}
			As mentioned in the Introduction, Horndeski theories are the most general four-dimensional, diffeomorphism covariant theories of gravity coupled to a scalar field which have second order equations of motion \cite{Horndeski1974}. The dynamical fields of such theories are a scalar field \(\Phi\) and the metric \(g\). The equations of motion can be obtained by varying the following action
			\begin{equation}
				S= \frac{1}{16\pi G}\int d^4 x \sqrt{-g} (\mathcal{L}_{1}+\mathcal{L}_{2}+\mathcal{L}_{3}+\mathcal{L}_{4}+\mathcal{L}_{5})
			\end{equation}
			where 
			\begin{align}
				\LL_1&=R + X - V(\Phi)\\
				\LL_{2}&= \GG_{2}(\Phi,X)\\
				\LL_{3}&= \GG_{3}(\Phi,X) \square\Phi\\
				\LL_{4}&= \GG_{4}(\Phi,X)R+\partial_{X}\GG_{4}(\Phi,X)\delta^{ac}_{bd}\,\nabla_{a}\nabla^{b}\Phi \, \nabla_{c}\nabla^{d}\Phi\\
			 	\LL_{5}&= \GG_{5}(\Phi,X)G_{ab}\nabla^{a}\nabla^{b}\Phi-\frac{1}{6}\partial_{X}\GG_{5}(\Phi,X)\delta^{ace}_{bdf}\,\nabla_{a}\nabla^{b}\Phi \, \nabla_{c}\nabla^{d}\Phi \, \nabla_{e}\nabla^{f}\Phi
			\end{align}
			and we have defined \(X=-\frac{1}{2}(\nabla \Phi)^{2}\). The functions \(\GG_{i}(\Phi,X)\) are arbitrary. We refer the reader to Appendix~\ref{sec:equations_of_motion_for_the_most_general_horndeski_theory} for the general form of the equations of motion obtained by varying this action.

			The first term, \(\LL_{1}\), corresponds to the standard Einstein--scalar-field theory, i.e. gravity minimally coupled to a scalar field \(\Phi\) with potential \(V(\Phi)\). We will assume that the functions \(\GG_{i}\) be smooth in their arguments.

			In order to study the hyperbolicity of these theories we will need to look at the linearised equations for a perturbation \((g+h,\Phi+\psi)\) of a background \((g,\Phi)\). 
			Moreover, since these theories are diffeomorphism-covariant, we will need to impose an appropriate gauge condition. 
			Following Ref.~\cite{Papallo2017}, we will choose the \emph{generalised
			harmonic gauge}.\footnote{This gauge can always be achieved via a gauge transformation and furthermore is propagated by the equations of motion. See Ref.~\cite{Papallo2017} for the details.}
			
			To obtain the linearised, generalised harmonic gauge Horndeski equations, we first expand the action to quadratic order in the perturbation \((h,\psi)\) and then we add to it the gauge fixing term
			\begin{equation}
				S_{\rm gauge}=-\frac{1}{2}\int \sqrt{- g} \, H_{a}H^{a} 
			\end{equation}
			where
			\begin{equation}
				\label{eq:ghg}
				H_{a}\equiv (1+f) G_a{}^{bcd} \nabla_b h_{cd} -\mathcal{H}_{a}{}^{b}\nabla_{b}\psi=0
			\end{equation}
			and \(\HH_{ab}\) and \(f\) depend only on background quantities. Moreover,
			they are assumed to be ``small'' and hence we can consider this gauge as a
			deformation of the standard harmonic gauge condition. Intuitively, as we
			deform the theory away from Einstein--scalar-field theory, 
			we might need to deform the gauge condition in order to preserve hyperbolicity. 
			Note that we could also add terms not involving derivatives of \(h_{ab}\) or 
			\(\psi\) as these would not modify the structure of the principal part of 
			\eqref{eq:ghg} and hence could not affect the hyperbolicity of the equations.
			The linearised gauge-fixed equations are then obtained by varying this action with respect to \(h\) and \(\psi\), and take the form
			\begin{align}
				P_{gg}^{abcdef}\nabla_{e}\nabla_{f}h_{cd}+P_{g\Phi}^{abef}\nabla_{e}\nabla_{f}\psi+\ldots  &  = 0\\
				P_{\Phi g}^{cdef}\nabla_{e}\nabla_{f}h_{cd}+P_{\Phi\Phi}^{ef}\nabla_{e}\nabla_{f}\psi+\ldots  & = 0
			\end{align}
			where the ellipses denotes terms with fewer than two derivatives. We can then
			read off the principal symbol for this system
			\begin{equation}
				P(\xi)=\begin{pmatrix}
					P_{gg}^{abcdef}\xi_{e}\xi_{f} & P_{g\Phi}^{abef}\xi_{e}\xi_{f}\\
					P_{\Phi g}^{cdef}\xi_{e}\xi_{f} & P_{\Phi\Phi}^{ef}\xi_{e}\xi_{f}
				\end{pmatrix}.
			\end{equation}
			We think of it as acting on vectors of the form \(T=(t_{ab},\chi)^{T}\), where \(t_{ab}\) is a symmetric 2-tensor and \(\chi\) is a scalar.
			It is convenient to split the principal symbol in its Einstein--scalar-field and Horndeski parts,
			\begin{equation}
				P(\xi)=P_{\rm Einstein}(\xi)+\delta P(\xi)
			\end{equation}
			where 
			\begin{equation}
				P_{\rm Einstein}(\xi)=
				\begin{pmatrix}
					-\frac{1}{2}\xi^{2}G^{abcd} & 0\\
					0 & -\xi^{2}
				\end{pmatrix}
			\end{equation}
			is the principal symbol for the harmonic gauge Einstein--scalar-field equations of motion and
			\begin{equation}
				G^{abcd}=\frac{1}{2}(g^{ac}g^{bd}+g^{ad}g^{bc}-g^{ab}g^{cd}).
			\end{equation}
			From this, it is easy to see that all characteristics of the generalised harmonic gauge Einstein--scalar-field system are null. We write
			\begin{equation}
				\delta P(\xi) = \delta \tilde{P}(\xi) + \delta Q(\xi)
			\end{equation}
			where \(\delta \tilde{P}\) denotes the terms arising from the Horndeski terms \(\LL_2, \LL_3,\LL_4, \LL_5\) in the action, and \(\delta Q\) denotes the \(f\) and \(\HH\)-dependent parts of the gauge-fixing terms. Explicitly we have
			\begin{equation}
				\delta Q(\xi)=
				\begin{pmatrix}
					-f(f+2) G^{abeh}G_{h}{}^{fcd}\xi_{e}\xi_{f} & (1+f)\xi^{e}G^{fhab}\xi_{h}\mathcal{H}_{ef}\\
					(1+f)\xi^{e}G^{fhcd}\xi_{h}\mathcal{H}_{ef} & -\mathcal{H}_{h}{}^{e}\mathcal{H}^{hf}\xi_{e}\xi_{f}
				\end{pmatrix}.
			\end{equation}
			Finally, we will be considering ``weak field'' backgrounds; i.e. we will
			assume that the background fields \((g,\Phi)\) are such that the Horndeski
			terms in the principal symbol 
			(see Appendix~\ref{sec:principal_symbol_for_the_most_general_horndeski_theory} 
			for its form) are small compared to those arising from Einstein--scalar-field
			theory  (i.e. all terms in Eqs.~\eqref{eq:dPgg_general_horndeski}--\eqref{eq:dPPhiPhi_general_horndeski} arising from \(\LL_{2}\)--\(\LL_{5}\) must be small compared to those arising from \(\LL_{1}\)).
			% That is to say we want all the terms in Eqs.~\eqref{eq:dPgg_general_horndeski}-\eqref{eq:dPPhiPhi_general_horndeski} to be small. 
			This is achieved by requiring the existence of an orthonormal basis in which 
			\begin{align}
				&|\GG_{4}-2X \partial_{X}\GG_{4}+X\partial_{\Phi}\GG_{5}|\ll1\\
				&|\partial_{X}\GG_{4}-\partial_{\Phi}\GG_{5}|L^{-2}\ll1\\
				&|X\partial_{X}\GG_{5}|L^{-2}  \ll1\\
				&\ldots\nonumber
			\end{align}
			and so on; we define \(L^{-2}\) as being the magnitude of the largest component of the Riemann curvature
            tensor, \((\nabla\Phi)^{2}\) or \(\nabla\nabla \Phi\) in this
            orthonormal basis.\footnote{Note that these conditions are weaker
            than those in Ref.~\cite{Papallo2017}. If the conditions in 
            Ref.~\cite{Papallo2017} hold, then these hold as well.}
			We will also require smallness of the functions appearing in the gauge condition:
			\begin{equation}
				\label{eq:gauge_small}
				|f| \ll 1, \qquad |{\cal H}_\mu{}^\nu| \ll 1 .
			\end{equation}
			% subsection equations_of_motion (end)
		\subsection{Weak and strong hyperbolicity} % (fold)
			\label{sub:weak_strong_hyperbolicity}
			Weak hyperbolicity is, roughly speaking, the statement that no solution to
			the equations of motion will grow exponentially fast in time, with the
			exponent proportional to the ``spatial frequency'' of the solution. Equivalently, we can think of it as the statement that any characteristic covector with real spatial components has a real time component. 
			In Ref.~\cite{Papallo2017} it was shown that 
			\medskip

			\noindent \emph{Horndeski theories are weakly hyperbolic around a generic weak field background, for any choice of generalised harmonic gauge.}
			\medskip

			Strong hyperbolicity refers to the existence of an ``energy estimate'' bounding an appropriate norm (in the case of a second order system, the \(H^{1}\)-norm) of the solution at some time \(t\) in terms of the initial data.
			Such inequality allows us, by standard methods, to establish local well-posedness of the (linearised) Cauchy problem.
			In Ref.~\cite{Papallo2017} a necessary condition for strong hyperbolicity to
			hold was derived:
			\medskip

			\noindent\emph{a necessary condition for strong hyperbolicity is that, for any null \(\xi\), \(\ker \delta P (\xi)\) has dimension greater or equal to 8.}
			\medskip

			It can be shown that for these theories, thanks to the symmetries of the principal symbol, we always have at least four elements in \(\ker \delta P(\xi)\) (for \(\xi\) null), namely the ``pure gauge'' vectors
			\begin{align}
				T=(\xi_{(a}Y_{b)}, 0)^{T},
			\end{align} 
			which arise from the residual gauge symmetry of the equations,
			\begin{align}
				h_{ab} \rightarrow h_{ab}+\nabla_{(a}Y_{b)}\qquad 
				\psi \rightarrow \psi+Y\cdot \nabla\Phi,
			\end{align}
			where \(Y_{a}\) satisfies \cite{Papallo2017}
			\begin{equation}
				\nabla^{b}\nabla_{b}Y_{a}- \frac{2}{1+f} \mathcal{H}_{a}{}^{b}\nabla_{b}(Y\cdot\nabla\Phi) 
				+ R_{ab} Y^{b}=0.
				%-\frac{2}{1+f} H_a \,.
			\end{equation}
			We can thus rephrase the condition stated above: for strong hyperbolicity to hold, it is necessary that, for null \(\xi\), \(\ker \delta P (\xi)\) contains at least four \emph{non-gauge} elements.
			In the rest of the paper, we will make use of this criterion to prove that some theories are not strongly hyperbolic.
			\medskip

			In Ref.~\cite{Papallo2017} it was proved that for \(\partial_X \GG_{4}\neq
			0\) and \(\GG_{5}=0\), there exists no generalised harmonic gauge in which the linearised Horndeski equations are strongly hyperbolic in a generic ``weak field'' background. On the other hand, we proved that when \(\partial_X \GG_{4}=\GG_{5}=0\), then there exists a unique generalised harmonic gauge,which makes the linearised Horndeski equations strongly hyperbolic in a generic ``weak field'' background. We also concluded that, by requiring the linearised generalised harmonic gauge condition to arise as the linearisation of a corresponding generalised harmonic gauge condition for the non-linear theory, one could further restrict the class of theories for which strong hyperbolicity holds to those with \(\GG_{3}=\partial_X \GG_{4}=\GG_{5}=0\):\footnote{Note that \(\GG_{4}\) can be made to vanish by an appropriate field redefinition.}
			\begin{equation}
				\label{eq:L_star}
				\LL_{*}=(1+\GG_{4}(\Phi))R+X-V(\Phi)+\GG_{2}(\Phi,X)
			\end{equation}
			with the gauge choice
			\begin{align}
				f=-1+\sqrt{1+\GG_{4}}, \qquad
				\mathcal{H}_{ab}= \, 2 \, \partial_{\Phi}f \, g_{ab}.
			\end{align}
			What remains to be investigated in detail is what happens when \(\GG_{5}\neq 0\).
			% subsection strong_hyperbolicity (end)
		% section background (end)
	\section{Failure of strong hyperbolicity for EdGB gravity} % (fold)
		\label{sec:einstein_dilaton_gauss_bonnet_gravity}
		We will begin our study by considering Einstein--dilaton--Gauss-Bonnet (EdGB)
		gravity \cite{Kanti1995}, a special case of the most general Horndeski theory,
		which was not covered in the analysis of Ref.~\cite{Papallo2017}. 
		The action for this theory is given by
		\begin{equation}
		    S=\frac{1}{16\pi}\int \sqrt{-g}\left(R+X+F(\Phi)L_{\rm GB}\right)
		\end{equation}
		where \(F(\Phi)\) is a smooth function and
		\begin{equation}
		    L_{\rm GB}=\frac{1}{4}\delta^{c_{1}c_{2}c_{3}c_{4}}_{d_{1}d_{2}d_{3}d_{4}}R_{c_{1}c_{2}}{}^{d_{1}d_{2}}R_{c_{3}c_{4}}{}^{d_{3}d_{4}}.
		\end{equation}
		The equations of motion take the form\footnote{Note that in the metric equation
		of motion the term 
		\(-\frac{1}{8}F(\Phi)\delta^{ac_{1}c_{2}c_{3}c_{4}}_{bd_{1}d_{2}d_{3}d_{4}}
		R_{c_{1}c_{2}}{}^{d_{1}d_{2}}R_{c_{3}c_{4}}{}^{d_{3}d_{4}}\) is sometimes
		included. However, in \(d=4\), we have that 
		\(\delta^{ac_{1}c_{2}c_{3}c_{4}}_{bd_{1}d_{2}d_{3}d_{4}}=0\), and hence this term vanishes identically.}
		\begin{align}
		   	E^{a}{}_{b} & \equiv  G^{a}{}_{b}+(F''(\Phi)\nabla_{c}\Phi\nabla^{d}\Phi+F'(\Phi)\nabla_{c}\nabla^{d}\Phi)\delta^{a c c_{1} c_{2}}_{b d d_{1} d_{2}}R_{c_{1}c_{2}}{}^{d_{1}d_{2}}-\frac{1}{2}T^{(\Phi)}_{ab}=0\\
		    E_{\Phi} & \equiv  -\square\Phi-\frac{1}{4}F'(\Phi)\delta^{c_{1}c_{2}c_{3}c_{4}}_{d_{1}d_{2}d_{3}d_{4}}R_{c_{1}c_{2}}{}^{d_{1}d_{2}}R_{c_{3}c_{4}}{}^{d_{3}d_{4}}=0
		\end{align}
		where
		\begin{equation}
		    T^{(\Phi)}_{ab}=\nabla_{a}\Phi\nabla_{b}\Phi+g_{ab}X.
		\end{equation}
		This theory can be cast in Horndeski form with the following choice of \(\GG_{i}\) \cite{Kobayashi2011}:
		\begin{align}
			\GG_{2}(\Phi,X)&=8 X^2 F^{(4)}(\Phi ) (3-\log |X|)\\
			\GG_{3}(\Phi,X)&=-4 X F^{(3)}(\Phi ) (7-3 \log |X|)\\
			\GG_{4}(\Phi,X)&=4 X F''(\Phi )(2-\log |X|) \\
			\GG_{5}(\Phi,X)&=-4  F'(\Phi )\log |X|.
		\end{align}
		Note that, in this case, while the functions \(\GG_{i}\) are not smooth at \(X=0\), the combinations that appear in the equations of motion (and in the principal symbol) are.
		We can linearise the generalised harmonic gauge EdGB equations of motion around a background \((g, \Phi)\) and compute the principal symbol. The part of the principal symbol which arises from the Horndeski terms reads
		\begin{align}
		    (\delta \tilde{P}_{gg}(\xi)\cdot t)^{a}{}_{b}& =  -2(F''(\Phi)\nabla_{c}\Phi\nabla^{d}\Phi+F'(\Phi)\nabla_{c}\nabla^{d}\Phi)\delta^{a c c_{1} c_{2}}_{b d d_{1} d_{2}}\xi_{c_{1}}\xi^{d_{1}}t_{c_{2}}{}^{d_{2}}\\
		    \delta\tilde{P}_{g\Phi}(\xi)^{a}{}_{b}=\delta\tilde{P}_{\Phi g}(\xi)^{a}{}_{b}& = F'(\Phi) \delta^{a c c_{1} c_{2}}_{b d d_{1} d_{2}}\xi_{c}\xi^{d}R_{c_{1}c_{2}}{}^{d_{1}d_{2}}\\
		    \delta \tilde{P}_{\Phi\Phi}(\xi)& =  0.
		\end{align}
		Note that for this theory the ``weak field'' conditions reduce to
		\begin{equation}
			|F'(\Phi)| L^{-2}\ll 1\qquad |F''(\Phi)| L^{-2}\ll 1.
		\end{equation}
		Following our discussion in Sec.~\ref{sub:weak_strong_hyperbolicity} we will study \(\ker \delta P(\xi)\), for null \(\xi\), to determine whether this theory can be strongly hyperbolic. 
		Recall that a vector \((t_{ab},\chi)^{T}\) is in \(\ker \delta P(\xi)\) if, and only if,
		\begin{equation}
			\label{eq:kernel_dP}
			\begin{pmatrix}
				\delta P_{gg}(\xi)^{abcd}t_{cd}+\delta P_{g\Phi}(\xi)^{ab}\chi\\
				\delta P_{\Phi g}(\xi)^{cd}t_{cd}+\delta P_{\Phi\Phi}(\xi)\chi
			\end{pmatrix}
			=0 .
		\end{equation}
		Let \(\xi\) be a null covector and look at the first row of \eqref{eq:kernel_dP},
		\begin{equation}
			\label{eq:first_line_ker_dP_edgb}
			\delta P_{gg}(\xi)^{abcd}t_{cd}+\delta P_{g\Phi}(\xi)^{ab}\chi=0.
		\end{equation}
		If this equation admitted a solution, then it could be used to fix \(t_{ab}\) 
		as a function of \(\chi\), up to the addition of linear combinations of
		elements of \(\ker \delta P_{gg}(\xi)\).
		It follows from the symmetries of the principal symbol that ``pure gauge'' vectors belong to this kernel \cite{Papallo2017}. If these were the only elements, then the above equation would fix completely the non-gauge part of \(t_{ab}\), implying that \(\ker\delta P(\xi)\) will contain at most one non-gauge element (and thus five elements at most), violating the necessary condition for strong hyperbolicity. 
		Hence, for strong hyperbolicity to hold, it is necessary that \(\ker \delta P_{gg}(\xi)\) contains non-gauge elements.
		We will therefore proceed to calculate \(\ker \delta P_{gg}(\xi)\).
		Defining, for convenience, the tensor
		\begin{equation}
			\FF_{ab}=-2(F''(\Phi)\nabla_{a}\Phi\nabla_{b}\Phi+F'(\Phi)\nabla_{a}\nabla_{b}\Phi)
		\end{equation}
		we can rewrite the condition for a tensor \(r_{ab}\) to be in \(\ker \delta P_{gg}(\xi)\), i.e. \(\delta P_{gg}(\xi)^{abcd}r_{cd}=0\), as
		\begin{equation}
		\label{eq:edgb_dPgg_kernel}
			\delta^{a c c_{1} c_{2}}_{b d d_{1} d_{2}}\xi_{c_{1}}\xi^{d_{1}}r_{c_{2}}{}^{d_{2}}\FF_{c}{}^{d}-\frac{1}{2}f(f+2)\delta^{ac_{1} c_{2}}_{b d_{1} d_{2}}\xi_{c_{1}}\xi^{d_{1}}r_{c_{2}}{}^{d_{2}}=0.
		\end{equation}
		In order to find solutions to this equation, we fix a point in spacetime and introduce a null basis \(\{e_{0},e_{1},e_{i}\}\) for the tangent space at that point. We take this basis to be adapted to \(\xi\), i.e. \(\xi^{\mu}=e_{0}^{\mu}\), \(e_{0}\cdot e_{0}=e_{1}\cdot e_{1}=0\), \(e_{0}\cdot e_{1}=1\), \(e_{i}\cdot e_{j}=\delta_{ij}\), and \(e_{0}\cdot e_{i}=e_{1}\cdot e_{i}=0\). In this basis, the system \((\delta P_{gg}(\xi)\cdot r)_{\mu\nu}=0\) reduces to
		\begin{align}
			\label{eq:EDGB_system}
			\frac{1}{2}f(f+2)(r_{22}+r_{33})-\FF_{33}r_{22}+2\FF_{23}r_{23}-\FF_{22}r_{33}&=0\\
			-\frac{1}{2}f(f+2)r_{02}+\FF_{33}r_{02}-\FF_{23}r_{03}-\FF_{03}r_{23}+\FF_{02}r_{33}&=0\\
			-\frac{1}{2}f(f+2)r_{03}+\FF_{22}r_{03}-\FF_{23}r_{02}+\FF_{03}r_{22}-\FF_{02}r_{23} &=0\\
			\frac{1}{2}f(f+2)r_{00}-\FF_{33}r_{00}+2\FF_{03}r_{03}-\FF_{00}r_{33} &=0\\
			\frac{1}{2}f(f+2)r_{00}-\FF_{22}r_{00}+2\FF_{02}r_{02}-\FF_{00}r_{22} &=0\\
			\label{eq:EDGB_system_last}
			\FF_{23}r_{00}-\FF_{03}r_{02}-\FF_{02}r_{03}+\FF_{00}r_{23} &=0.
		\end{align}
		Note that the ``gauge'' components of \(r_{ab}\), i.e., \(r_{1\mu}\), do not appear in the equations.
		This is a system of six linear equations for the six non-gauge components of \(r_{ab}\). Since the number of unknowns equals the number of equations, generically, this system  will have no non-trivial solution unless the determinant of the matrix of coefficients vanishes. We will now show that this determinant does not vanish for any choice of generalised Harmonic gauge.

		The matrix of coefficients takes the form
		\begin{equation}
			C=\begin{pmatrix}
				0 & 0 & 0 & \frac{f(f+2)}{2}-\FF_{33} & 2 \FF_{23} & \frac{f(f+2)}{2}-\FF_{22} \\
 				0 & \FF_{33}-\frac{f(f+2)}{2} & -\FF_{23} & 0 & -\FF_{03} & \FF_{02} \\
				0 & -\FF_{23} & \FF_{22}-\frac{f(f+2)}{2} & \FF_{03} & -\FF_{02} & 0 \\
				\frac{f(f+2)}{2}-\FF_{33} & 0 & 2 \FF_{03} & 0 & 0 & -\FF_{00} \\
				\FF_{23} & -\FF_{03} & -\FF_{02} & 0 & \FF_{00} & 0 \\
				\frac{f(f+2)}{2}-\FF_{22} & 2 \FF_{02} & 0 & -\FF_{00} & 0 & 0 \\
			\end{pmatrix}.
		\end{equation}
		In the null basis, its determinant reads
		\begin{align}
			\det C =-\frac{1}{8}\,\biggl[ & \FF_{00} (f(f+2))^2+f(f+2) \left(-2 \FF_{00} \FF_{22}-2 \FF_{00} \FF_{33}+2 \FF_{02}^2+2 \FF_{03}^2\right)\nonumber\\
			&+4 \FF_{00}\FF_{22} \FF_{33}-4 \FF_{00} \FF_{23}^2-4 \FF_{02}^2 \FF_{33}+8 \FF_{02} \FF_{03} \FF_{23}-4 \FF_{03}^2 \FF_{22} \biggr]^{2}.
		\end{align}
		The condition \(\det C=0\) can be rewritten covariantly and is equivalent to the following
		\begin{align}
			\label{eq:det_system}
		 	f^{2}(f+2)^{2}\xi^{c}\xi^{d}\FF_{cd} -2f(f+2)(\delta^{c_{1}c_{2}c_{3}}_{d_{1}d_{2}d_{3}}\xi_{c_{1}}\xi^{d_{1}}\FF_{c_{2}}{}^{d_{2}}\FF_{c_{3}}{}^{d_{3}})-\frac{8}{3}\delta^{c_{1}c_{2}c_{3}c_{4}}_{d_{1}d_{2}d_{3}d_{4}}\xi_{c_{1}}\xi^{d_{1}}\FF_{c_{2}}{}^{d_{2}}\FF_{c_{3}}{}^{d_{3}}\FF_{c_{4}}{}^{d_{4}}=0.
		\end{align} 
		Solving for \(f\) gives \(f=f_{*}\), with\footnote{Note that there is only one choice of sign in front of the square root for which \(f_{*}\) satisfies the smallness assumptions \eqref{eq:gauge_small}.}
		\begin{align}
			f_{*}=-1+\sqrt{(1-2\FF^{e}{}_{e})+\frac{2\FF_{ae}\FF_{b}{}^{e}\xi^{a}\xi^{b}\pm\sqrt{A}}{\FF_{cd}\xi^{c}\xi^{d}}}
		\end{align}
		where
		\begin{equation}
			A=(\delta^{c_{1}c_{2}c_{3}}_{d_{1}d_{2}d_{3}}\xi_{c_{1}}\xi^{d_{1}}\FF_{c_{2}}{}^{d_{2}}\FF_{c_{3}}{}^{d_{3}})^{2}+\frac{8}{3}\FF_{b}{}^{a}\xi_{a}\xi^{b}\delta^{c_{1}c_{2}c_{3}c_{4}}_{d_{1}d_{2}d_{3}d_{4}}\xi_{c_{1}}\xi^{d_{1}}\FF_{c_{2}}{}^{d_{2}}\FF_{c_{3}}{}^{d_{3}}\FF_{c_{4}}{}^{d_{4}}.
		\end{equation}
		Note that \(f\) should only depend on background fields, it cannot depend on \(\xi\).
		The function \(f_{*}\) could only be independent of \(\xi\) if the second term in the square root in the above expression were independent of \(\xi\); that is if 
		\begin{equation}
			2\FF_{ae}\FF_{b}{}^{e}\xi^{a}\xi^{b}\pm\sqrt{A}=\lambda \FF_{ab}\xi^{a}\xi^{b}
		\end{equation}
		for some scalar \(\lambda\) independent of \(\xi\). By rearranging the terms and squaring them, we see that we can equivalently look for a \(\lambda\) which solves (expanding \(A\) and using \(\xi^{2}=0\))
		\begin{equation}
		\label{eq:lambda_eq}
			\FF_{ab}\xi^{a}\xi^{b}\left[\FF_{cd}\xi^{c}\xi^{d}(\lambda^{2}-8\FF_{ef}\FF^{ef}+4\FF^{e}{}_{e}\FF^{f}{}_{f})-4\FF_{c}{}^{e}\FF_{de}\xi^{c}\xi^{d}(\lambda+2\FF^{e}{}_{e})+16 \FF^{ef}\FF_{ce}\FF_{df}\xi^{c}\xi^{d}\right]=0.
		\end{equation}
		However, since the three terms in square parenthesis have different dependence on \(\xi\), they would have to vanish independently, and there is no choice of \(\lambda\) for which this happens in a generic background. Note that there is no special choice of \(F(\Phi)\) for which this result would be different. To see this, we substitute the explicit form of \(\FF_{ab}\) in Eq.~\eqref{eq:lambda_eq}; the last term will give rise to a term of the form 
		\begin{equation}
		\label{eq:lambda_argument}
			(F'(\Phi))^{3}\nabla^{e}\nabla^{f}\Phi\nabla_{c}\nabla_{e}\Phi\nabla_{d}\nabla_{f}\Phi \xi^{c}\xi^{d}.
		\end{equation}
		Since there is no other term involving \(\xi\) contracted with the same
		combination of derivatives of \(\Phi\), for \eqref{eq:lambda_eq} to hold for \(\lambda\) independent of \(\xi\), it is necessary that this term vanishes in a generic background, i.e., \(F'(\Phi)=0\). However, for such choice of \(F(\Phi)\), EdGB theory would reduce to GR.

		We can deduce from this argument that that any \(f_{*}\) which solves \(\det C=0\) would necessarily depend on \(\xi\), which is not allowed. This implies that for any ``good'' choice of the function \(f\), in a generic background (for which \(\FF_{ab}\) does not have any special properties) the system \eqref{eq:EDGB_system}--\eqref{eq:EDGB_system_last} has no non-trivial solution.
		Therefore we can conclude that, in a generic weak field background, for any choice of generalised harmonic gauge, the only elements of \(\ker \delta P_{gg}(\xi)\) are ``pure gauge'' vectors, i.e., \(r_{ab}=\xi_{(a}X_{b)}\). 

		Going back to our original question, this result implies that if a \(t_{ab}\) solving \eqref{eq:first_line_ker_dP_edgb} exists, then this solution will be unique up to the addition of multiples of ``pure gauge'' vectors; 
		that is Eq.~\eqref{eq:first_line_ker_dP_edgb} completely fixes the non-gauge
		part of \(t_{ab}\) in terms of \(\chi\). If we then substituted such \(t_{ab}\) in the second row of \eqref{eq:kernel_dP}, we would obtain a linear homogeneous equation for \(\chi\) which, for a generic background, would only admit the solution \(\chi=0\). This would in turn imply that the ``non-gauge'' part of \(t_{ab}\) had to vanish, i.e., \(t_{ab}=\xi_{(a}Y_{b)}\). Hence, we can conclude that, for any choice of generalised harmonic gauge, in a generic ``weak field'' background we have that \(\dim\ker\delta P(\xi)=4\) and thus the linearised EdGB equations are not strongly hyperbolic in this setting.
	\section{Failure of strong hyperbolicity for a general Horndeski theory} % (fold)
		\label{sec:failure_of_strong_hyperbolicity_for_a_general_horndeski_theory}
		We will now investigate whether general Horndeski theories are strongly
		hyperbolic or they suffer from similar problems. In Ref.~\cite{Papallo2017}
		the hyperbolicity of general Horndeski theories with \(\GG_{5}=0\) was studied in detail. It was found that the most general such theory with strongly hyperbolic equations of motion (in a generalised harmonic gauge and for a generic ``weak field'' background) was the one obtained by setting \(\partial_X\GG_{4}=\GG_{3}=0\), i.e. that corresponding to the Lagrangian \(\LL_{*}\) (Eq.~\eqref{eq:L_star}). In this section we will study theories with \(\GG_{5}\neq 0\).

		\subsection{\texorpdfstring{\(\GG_{5}=\GG_{5}(\Phi)\)}{G5Phi} case}
		Consider first the case in which \(\GG_{5}=\GG_{5}(\Phi)\) already discussed
		in Ref.~\cite{Papallo2017}. The corresponding contribution to the Horndeski
		Lagrangian will read
		\begin{equation}
			\LL_{5}=\GG_{5}(\Phi)G_{ab}\nabla^{a}\nabla^{b}\Phi.
		\end{equation}
		However, it can be shown that this is equivalent -- up to a total derivative term, which does not contribute to the equations of motion -- to \cite{Kobayashi2011,Bettoni2013}
		\begin{align}
			\LL_{5}=-\partial_{\Phi}\GG_{5} X R -\partial_{\Phi}\GG_{5}\delta^{ac}_{bd}\nabla_{a}\nabla^{b}\Phi\nabla_{c}\nabla^{d}\Phi+3\partial^{2}_{\Phi}\GG_{5} X \square \Phi - 2 \partial^{3}_{\Phi}\GG_{5} X^{2},
		\end{align}
		and hence the full Horndeski Lagrangian can be rewritten as 
		\begin{equation}
			\LL=\LL_{1}+\tilde{\LL}_{2}+\tilde{\LL}_{3}+\tilde{\LL}_{4},
		\end{equation}
		where
		\begin{align}
			\tilde{\GG}_{2}=\GG_{2}-2\partial^{3}_{\Phi}\GG_{5}X^{2}\qquad
			\tilde{\GG}_{3}=\GG_{3}+3\partial^{2}_{\Phi}\GG_{5}X\qquad
			\tilde{\GG}_{4}=\GG_{4}-\partial_{\Phi}\GG_{5}X.
		\end{align}
		As this is effectively equivalent to a Horndeski theory with \(\tilde{\GG}_{5}=0\), it will be strongly hyperbolic around a generic ``weak field'' background, in a generalised harmonic gauge if, and only if, 
		\linebreak
		\(\partial_{X}\tilde{\GG}_{4}=0=\tilde{\GG}_{3}\), i.e., iff
		\begin{align}
			\GG_{3}=-3\partial^{2}_{\Phi}\GG_{5}X \qquad \partial_{X}\GG_{4}=\partial_{\Phi}\GG_{5}.
		\end{align}
		However, with this choice, the theory simply reduces to \(\LL_{*}\). In order to avoid this degeneracy, we will need \(\GG_{5}\) to depend on \(X\).
		% \medskip
		\subsection{\texorpdfstring{\(\GG_{5}=\GG_{5}(\Phi,X)\)}{G5XPhi} case}

		Let us consider the most general Horndeski theory, i.e., \(\GG_{i}\neq 0\) and
		\(\GG_{5}=\GG_{5}(\Phi,X)\neq 0\). The equations of motion and the principal
		symbol for this theory are reported in Appendix~\ref{sec:equations_of_motion_for_the_most_general_horndeski_theory} and \ref{sec:principal_symbol_for_the_most_general_horndeski_theory}, respectively. In Section~IV.D.4 of Ref.~\cite{Papallo2017} it was argued that given the structure of the equations of motion, one would not expect such theory to be strongly hyperbolic. We will now provide a rigorous proof of this claim, based on the results of Sec.~\ref{sec:einstein_dilaton_gauss_bonnet_gravity}. 

		The main obstruction to strong hyperbolicity in EdGB theory, as discussed in the previous section, arose from the fact that the dimension of the kernel of \(\delta P_{gg}(\xi)\) was not large enough. We will therefore proceed to study the corresponding operator in the general Horndeski theory. From Eq.~\eqref{eq:dPgg_general_horndeski}, we see that, for null \(\xi\), it takes the form
		\begin{align}
		 	(\delta P_{gg}(\xi)\cdot t)^{a}{}_{b}&= 
		 	\left(F_{1}(\Phi,X)-\frac{1}{2}f(f+2)\right)
            \delta^{a c_{1}c_{2}}_{b d_{1}d_{2}}\xi_{c_{1}}\xi^{d_{1}}t_{c_{2}}{}^{d_{2}}\nonumber\\
		 	&+F_{2}(\Phi,X)\delta^{a c_{1} c_{2} c_{3}}_{b d_{1} d_{2} d_{3}}
            \xi_{c_{1}}\xi^{d_{1}}t_{c_{2}}{}^{d_{2}}\nabla_{c_{3}}\Phi\nabla^{d_{3}}\Phi\nonumber\\
		 	&+F_{3}(\Phi,X)\delta^{a c_{1} c_{2} c_{3}}_{b d_{1} d_{2} d_{3}}
            \xi_{c_{1}}\xi^{d_{1}}t_{c_{2}}{}^{d_{2}}\nabla_{c_{3}}\nabla^{d_{3}}\Phi
		 \end{align} 
		where
		\begin{align}
		 	F_{1}(\Phi,X)&=-\frac{1}{2}(\GG_{4}-2X \partial_{X}\GG_{4}+X \partial_{\Phi}\GG_{5})\\
		 	F_{2}(\Phi,X)&=-\frac{1}{2}(\partial_{X}\GG_{4}- \partial_{\Phi}\GG_{5})\\
		 	F_{3}(\Phi,X)&=\frac{1}{2}X \partial_{X}\GG_{5}.
		\end{align}
		We will assume \(F_{3}\neq 0\), for otherwise the theory would reduce to the case \(\GG_{5}=\GG_{5}(\Phi)\) discussed earlier.
		Consider now the condition \eqref{eq:kernel_dP} for a vector \(T=(t_{ab},\chi)^{T}\) to belong to \(\ker \delta P(\xi)\). The first row reads
		\begin{equation}
			\label{eq:first_line_ker_dP}
			\delta P_{gg}(\xi)^{abcd}t_{cd}+\delta P_{g\Phi}(\xi)\chi=0.
		\end{equation}
		In the EdGB case, we used the fact that \(\ker \delta P_{gg}(\xi)\) only contained ``pure gauge'' elements to conclude that if this equation admitted a solution then it would fix completely the ``non-gauge'' part of \(t_{ab}\) in terms of \(\chi\), implying that the dimension of \(\ker \delta P(\xi)\) could not be large enough for strong hyperbolicity to hold. We will now show that the same statement holds in the general case. Consider the equation 
		\begin{equation}
			\label{eq:ker_dPgg_horndeski}
			(\delta P_{gg}(\xi)\cdot r)_{ab}=0.
		\end{equation}
		Its tensorial structure is essentially identical to that of the corresponding equation in EdGB theory, Eq.~\eqref{eq:edgb_dPgg_kernel}.
		In fact, defining 
		\begin{align}
			\tilde{\FF}_{ab}&=F_{2}(\Phi,X)\nabla_{a}\Phi\nabla_{b}\Phi+F_{3}(\Phi,X)\nabla_{a}\nabla_{b}\Phi\\
			\tilde{f}&=-1+\sqrt{(1+f)^{2}-2F_{1}}
		\end{align}
		Eq.~\eqref{eq:ker_dPgg_horndeski} takes the form
		\begin{equation}
			\label{eq:ker_dPgg_horndeski_2}
			\delta^{a c c_{1} c_{2}}_{b d d_{1} d_{2}}\xi_{c_{1}}\xi^{d_{1}}r_{c_{2}}{}^{d_{2}}\tilde{\FF}_{c}{}^{d}-\frac{1}{2}\tilde{f}(\tilde{f}+2)\delta^{ac_{1} c_{2}}_{b d_{1} d_{2}}\xi_{c_{1}}\xi^{d_{1}}r_{c_{2}}{}^{d_{2}}=0.
		\end{equation}
		Note that even if \(F_{1}=0\) or \(F_{2}=0\), the form of this equation would be unchanged and hence we do not need to consider these cases separately.
		We can study this system in the same way as we studied \eqref{eq:EDGB_system}--\eqref{eq:EDGB_system_last}. Recall that this is a system of six equations for the six ``non-gauge'' components of \(r_{ab}\) (the ``pure gauge'' components do not appear in these equations). An analogous argument allows us to conclude that there is no admissible choice of \(f\) for which this system would generically admit non-trivial solutions.\footnote{In this setting, Eq.~\eqref{eq:lambda_eq} would have a solution independent of \(\xi\) if, and only if, \(F_{3}=0\). However, since we are assuming \(F_{3}\neq 0\), this does not happen. To see that this condition is necessary, we could repeat the argument following Eq.~\eqref{eq:lambda_eq} where now \(F_{3}\) plays the role of \(F'\) in Eq.~\eqref{eq:lambda_argument}.} Therefore, we can deduce that for a generic ``weak field'' background and for any choice of generalised harmonic gauge, \(\ker \delta P_{gg}(\xi)\) will only contain ``pure gauge'' elements. 

		Since \(\ker \delta P_{gg}(\xi)\) contains only ``pure gauge'' elements, then
		if a solution to \eqref{eq:first_line_ker_dP} exists, it will be unique up to
		addition of multiples of ``pure gauge'' vectors. In other words, Eq.~\eqref{eq:first_line_ker_dP} completely fixes the non-gauge part of a \(t_{ab}\) in \(\ker \delta P (\xi)\) in terms of \(\chi\). If we then substitute such \(t_{ab}\) into the second row of \eqref{eq:kernel_dP}, we will obtain a linear homogeneous equation for \(\chi\), which for generic background has only trivial solutions \(\chi=0\). This implies that all elements in \(\ker \delta P(\xi)\) have vanishing ``non-gauge'' part, i.e., \(\ker \delta P(\xi)\) only contains ``pure gauge'' elements.

		Finally, since \(\dim \ker \delta P(\xi)=4<8\), we conclude that the necessary condition for strong hyperbolicity stated in Section~\ref{sec:background} is not satisfied and hence the most general Horndeski theory fails to be strongly hyperbolic in a generic ``weak field'' background for any choice of generalised harmonic gauge.
		% section failure_of_strong_hyperbolicity_for_a_general_horndeski_theory (end)	
		% \pagebreak
	\section{Discussion} % (fold)
		\label{sec:discussion}
		In this paper, we concluded the analysis of the hyperbolicity of Horndeski
		theories initiated in Ref.~\cite{Papallo2017}. 
		
		First, we considered the equations of motion of Einstein--dilaton--Gauss-Bonnet gravity linearised around a generic ``weak field'' background and we showed that these fail to be strongly hyperbolic in any generalised harmonic gauge. 

		We then studied the equations of motion for the most general Horndeski theory (with \(\GG_{5}\neq 0\)). Since the structure of the principal symbol turns out to be similar to that of EdGB, we were able to employ the results obtained earlier to conclude that there is no choice of generalised harmonic gauge for which these equations are strongly hyperbolic in a generic ``weak field'' background.

		We conclude that, as expected from the considerations in Ref.~\cite{Papallo2017}, the most general Horndeski theory which, in a generalised harmonic gauge, is strongly hyperbolic in a generic weak field background is that given by \(\LL_{*}\), that is to say, Einstein gravity coupled to a ``k-essence'' field (Eq.~\eqref{eq:L_star}).

		Recall that without strong hyperbolicity, the best one can hope to prove is that the linearised Cauchy problem be locally well-posed with a ``loss of derivatives''. However, even if this were the case, such loss of derivatives would constitute a serious obstruction to proving local well-posedness for the non-linear problem. For this reason, we do not expect any Horndeski theory more general than \(\LL_{*}\) to have a well-posed initial value problem in generalised harmonic gauge.

		Our results hold for the generalised harmonic gauge equations and so, in principle, there could be a different choice of gauge in which a larger subclass of Horndeski theories were strongly hyperbolic. Moreover, one may consider different methods to tackle the problem such as a (suitably modified) ADM-type decomposition or studying an evolution equation for the curvature tensor obtained from the Bianchi identity instead of the evolution equation for the metric. We refer the reader to the discussion in Ref.~\cite{Papallo2017} for a more exhaustive treatment of these alternative approaches to the problem and related issues. In addition to the remarks pointed out there, note that the way we obtained the generalised harmonic gauge equations is by no means unique. One could, for example, use different \(\HH_{ab}\) in the metric and scalar field gauge fixing terms. However, fixing the gauge in this way would break the symmetry of the principal symbol: \(P_{g\Phi}\neq P_{\Phi g}\). Our proof of weak hyperbolicity relies crucially on this symmetry and therefore it seems unlikely that such modification of the gauge fixing term could be useful.

		% \medskip
	% section discussion (end)
	\section*{Acknowledgements}
		% \ldots
		 % \paragraph{Acknowledgements}
		%  % \ldots
		I would like to thank my supervisor, Harvey Reall, for suggesting the problem, for the many useful discussions and for the comments on this paper.
		I am also grateful to Frans Pretorius for an interesting conversation on EdGB theory, and to Aron Kovacs for his feedback on an earlier version of this paper.
		This research was supported by an STFC studentship and a Cambridge
		Philosophical Society Research Grant.
	\appendix
	\section{Equations of motion for the most general Horndeski theory} % (fold)
		\label{sec:equations_of_motion_for_the_most_general_horndeski_theory}
		For reference, we include here the most general form of the Horndeski equations (i.e. when all \(\GG_{i}(\Phi,X)\neq 0\), \(i=2,3,4,5\)):
		\begin{align}
			\label{eq:horndeski_full_eq_metric}
		    E^{a}{}_{b}\equiv&-\frac{1}{4}\left[1+\GG_{4}-2X \partial_{X}\GG_{4}+X\partial_{\Phi}\GG_{5}\right]\delta^{a c_{1}c_{2}}_{b d_{1}d_{2}}R_{c_{1}c_{2}}{}^{d_{1}d_{2}}\nonumber\\
		    &+\frac{1}{4}\left[\partial_{X}\GG_{4}-\partial_{\Phi}\GG_{5}\right]\delta^{a c_{1}c_{2}c_{3}}_{b d_{1}d_{2}d_{3}}\nabla_{c_{1}}\Phi\nabla^{d_{1}}\Phi R_{c_{2}c_{3}}{}^{d_{2}d_{3}}\nonumber\\
		    &-\frac{1}{4}\left[X \partial_{X} \GG_{5}\right]\delta^{a c_{1}c_{2}c_{3}}_{b d_{1}d_{2}d_{3}}\nabla_{c_{1}}\nabla^{d_{1}}\Phi R_{c_{2}c_{3}}{}^{d_{2}d_{3}}\nonumber\\
		    &-\frac{1}{2}\left(2X+\GG_{2}+2X\partial_{\Phi}\GG_{3}+4X \partial^{2}_{\Phi}\GG_{4}\right)\delta^{a}_{b}-\frac{1}{2}\left(2+\partial_{X}\GG_{2}+2\partial_{\Phi}\GG_{3}+2\partial^{2}_{\Phi}\GG_{4}\right)\nabla^{a}\Phi\nabla_{b}\Phi\nonumber\\
		    &+\left[X\partial_{X}\GG_{3}+\partial_{\Phi}\GG_{4}+2X\partial^{2}_{X \Phi}\GG_{4}\right]\delta^{a c}_{b d}\nabla_{c}\nabla^{d}\Phi\nonumber\\
		    &+\frac{1}{2}\left[4\partial^{2}_{X\Phi}\GG_{4}+\partial_{X}\GG_{3}-\partial^{2}_{\Phi}\GG_{5}\right]\delta^{a c_{1}c_{2}}_{b d_{1}d_{2}}\nabla_{c_{1}}\Phi\nabla^{d_{1}}\Phi\nabla_{c_{2}}\nabla^{d_{2}}\Phi\nonumber\\
		    &+\frac{1}{2}\left[\partial_{X}\GG_{4}+2X \partial^{2}_{X}\GG_{4}-\partial_{\Phi}\GG_{5}-X\partial^{2}_{X \Phi}\GG_{5}\right]\delta^{a c_{1}c_{2}}_{b d_{1} d_{2}}\nabla_{c_{1}}\nabla^{d_{1}}\Phi\nabla_{c_{2}}\nabla^{d_{2}}\Phi\nonumber\\
		    &+\frac{1}{2}\left[\partial^{2}_{X}\GG_{4}-\partial^{2}_{X\Phi}\GG_{5}\right]\delta^{a c_{1}c_{2}c_{3}}_{b d_{1}d_{2}d_{3}}\nabla_{c_{1}}\nabla^{d_{1}}\Phi\nabla_{c_{2}}\nabla^{d_{2}}\Phi\nabla_{c_{3}}\Phi\nabla^{d_{3}}\Phi\nonumber\\
		    &-\frac{1}{6}[\partial_{X}\GG_{5}+X\partial^{2}_{X}\GG_{5}]\delta^{a c_{1}c_{2} c_{3}}_{b d_{1} d_{2}d_{3}}\nabla_{c_{1}}\nabla^{d_{1}}\Phi\nabla_{c_{2}}\nabla^{d_{2}}\Phi\nabla_{c_{3}}\nabla^{d_{3}}\Phi=0\\
		    \label{eq:horndeski_full_eq_scalar_field}
		    E_{\Phi}\equiv &-[1+\partial_{X}\GG_{2}+2X \partial^{2}_{X}\GG_{2}+2\partial_{\Phi}\GG_{3}+2X\partial^{2}_{X\Phi}\GG_{3}]\square\Phi\nonumber\\
		    &-[\partial^{2}_{X}\GG_{2}+2\partial^{2}_{X \Phi}\GG_{3}+2\partial^{3}_{X\Phi\Phi}\GG_{4}]\delta^{c_{1}c_{2}}_{d_{1}d_{2}}\nabla_{c_{1}}\Phi\nabla^{d_{1}}\Phi\nabla_{c_{2}}\nabla^{d_{2}}\Phi\nonumber\\
		    &-[\partial_{X}\GG_{3}+X\partial^{2}_{X}\GG_{3}+2X\partial^{3}_{XX\Phi}\GG_{4}+3\partial^{2}_{X\Phi}\GG_{4}]\delta^{c_{1}c_{2}}_{d_{1}d_{2}}\nabla_{c_{1}}\nabla^{d_{1}}\Phi\nabla_{c_{2}}\nabla^{d_{2}}\Phi\nonumber\\
		    &-\frac{1}{4}[\partial_{X}\GG_{3}+4\partial^{2}_{X\Phi}\GG_{4}-\partial^{2}_{\Phi}\GG_{5}]\delta^{c_{1}c_{2}c_{3}}_{d_{1} d_{2}d_{3}}\nabla_{c_{1}}\Phi\nabla^{d_{1}}\Phi R_{c_{2}c_{3}}{}^{d_{2}d_{3}}\nonumber\\
		    &-[X\partial_{X}\GG_{3}+\partial_{\Phi}\GG_{4}+2X\partial^{2}_{X\Phi}\GG_{4}]R\nonumber\\
		    &-\frac{1}{2}[\partial^{2}_{X}\GG_{3}+4\partial^{3}_{XX\Phi}\GG_{4}-\partial^{3}_{X\Phi\Phi}\GG_{5}]\delta^{c_{1}c_{2}c_{3}}_{d_{1}d_{2}d_{3}}\nabla_{c_{1}}\nabla^{d_{1}}\Phi\nabla_{c_{2}}\nabla^{d_{2}}\Phi\nabla_{c_{3}}\Phi\nabla^{d_{3}}\Phi\nonumber\\
		    &-\frac{1}{2}[\partial_{X}\GG_{4}+2X \partial^{2}_{X}\GG_{4}-\partial_{\Phi}\GG_{5}-X\partial^{2}_{X\Phi}\GG_{5}]\delta^{c_{1}c_{2}c_{3}}_{d_{1}d_{2}d_{3}}\nabla_{c_{1}}\nabla^{d_{1}}\Phi R_{c_{2}c_{3}}{}^{d_{2}d_{3}}\nonumber\\
		    &-\frac{1}{2}[\partial^{2}_{X}\GG_{4}-\partial^{2}_{X\Phi}\GG_{5}]\delta^{c_{1}c_{2}c_{3}c_{4}}_{d_{1}d_{2}d_{3}d_{4}}\nabla_{c_{1}}\nabla^{d_{1}}\Phi\nabla_{c_{2}}\Phi\nabla^{d_{2}}\Phi R_{c_{3}c_{4}}{}^{d_{3}d_{4}}\nonumber\\
		    &-\frac{1}{3}\left[3\partial^{2}_{X}\GG_{4}+2X\partial^{3}_{X}\GG_{4}-2\partial^{2}_{X\Phi}\GG_{5}-X\partial^{3}_{XX\Phi}\GG_{5}\right]\delta^{c_{1}c_{2}c_{3}}_{d_{1}d_{2}d_{3}}\nabla_{c_{1}}\nabla^{d_{1}}\Phi\nabla_{c_{2}}\nabla^{d_{2}}\Phi\nabla_{c_{3}}\nabla^{d_{3}}\Phi\nonumber\\
		    &-\frac{1}{3}[\partial^{3}_{X}\GG_{4}-\partial^{3}_{XX\Phi}\GG_{5}]\delta^{c_{1}c_{2}c_{3}c_{4}}_{d_{1}d_{2}d_{3}d_{4}}\nabla_{c_{1}}\nabla^{d_{1}}\Phi\nabla_{c_{2}}\nabla^{d_{2}}\Phi\nabla_{c_{3}}\nabla^{d_{3}}\Phi\nabla_{c_{4}}\Phi\nabla^{d_{4}}\Phi\nonumber\\
		    &+\frac{1}{12}[2\partial^{2}_{X}\GG_{5}+X\partial^{3}_{X}\GG_{5}]\delta^{c_{1}c_{2}c_{3}c_{4}}_{d_{1}d_{2}d_{3}d_{4}}\nabla_{c_{1}}\nabla^{d_{1}}\Phi\nabla_{c_{2}}\nabla^{d_{2}}\Phi\nabla_{c_{3}}\nabla^{d_{3}}\Phi\nabla_{c_{4}}\nabla^{d_{4}}\Phi\nonumber\\
		    &+\frac{1}{4}[\partial_{X}\GG_{5}+X\partial^{2}_{X}\GG_{5}]\delta^{c_{1}c_{2}c_{3}c_{4}}_{d_{1}d_{2}d_{3}d_{4}}\nabla_{c_{1}}\nabla^{d_{1}}\Phi\nabla_{c_{2}}\nabla^{d_{2}}\Phi R_{c_{3}c_{4}}{}^{d_{3}d_{4}}\nonumber\\
		    &+\frac{1}{16}X\partial_{X}\GG_{5}\delta^{c_{1}c_{2}c_{3}c_{4}}_{d_{1}d_{2}d_{3}d_{4}} R_{c_{1}c_{2}}{}^{d_{1}d_{2}} R_{c_{3}c_{4}}{}^{d_{3}d_{4}} \nonumber\\
		    &+2X(\partial^{2}_{\Phi}\GG_{3}+\partial^{2}_{X \Phi}\GG_{2})-\partial_{\Phi}\GG_{2}=0.
		\end{align}
	% section equations_of_motion_for_the_most_general_horndeski_theory (end)
	\section{Principal symbol for the most general Horndeski theory} % (fold)
		\label{sec:principal_symbol_for_the_most_general_horndeski_theory}
		By linearising the equations \eqref{eq:horndeski_full_eq_metric}--\eqref{eq:horndeski_full_eq_scalar_field} around a background \((g,\Phi)\), we can compute the principal symbol of the most general Horndeski theory.
		Recall that in Sec.~\ref{sec:background} we have separated the principal
		symbol into three parts: 
		\(\delta P(\xi)=P_{\rm Einstein}(\xi)+ \delta \tilde{P}(\xi)+\delta Q(\xi)\),
		where \(\delta P_{\rm Einstein}(\xi)\) is the principal symbol for the 
		(generalised) harmonic gauge Einstein equations, \(\delta Q(\xi)\) denotes the \(f\)- and \(\HH\)-dependent parts of the gauge-fixing terms and \(\delta \tilde{P}(\xi)\) denotes the contribution to the principal symbol arising from the Horndeski terms. We include here the explicit form of the latter for the most general Horndeski theory:
		\begin{align}
			\label{eq:dPgg_general_horndeski}
		    (\delta \tilde{P}_{gg}(\xi)\cdot t)^{a}{}_{b}=& 
		 	-\frac{1}{2}(\GG_{4}-2X \partial_{X}\GG_{4}+X \partial_{\Phi}\GG_{5})\delta^{a c_{1}c_{2}}_{b d_{1} d_{2}}\xi_{c_{1}}\xi^{d_{1}}t_{c_{2}}{}^{d_{2}}\nonumber\\
		 	&-\frac{1}{2}(\partial_{X}\GG_{4}- \partial_{\Phi}\GG_{5})\delta^{a c_{1} c_{2} c_{3}}_{b d_{1} d_{2} d_{3}}\xi_{c_{1}}\xi^{d_{1}}t_{c_{2}}{}^{d_{2}}\nabla_{c_{3}}\Phi\nabla^{d_{3}}\Phi\nonumber\\
		 	&+\frac{1}{2}X \partial_{X}\GG_{5}\delta^{a c_{1} c_{2} c_{3}}_{b d_{1} d_{2} d_{3}}\xi_{c_{1}}\xi^{d_{1}}t_{c_{2}}{}^{d_{2}}\nabla_{c_{3}}\nabla^{d_{3}}\Phi\\
		    \label{eq:dPgPhi_general_horndeski}
		    \delta \tilde{P}_{g\Phi}(\xi)^{a}{}_{b}=\delta \tilde{P}_{\Phi g}(\xi)^{a}{}_{b}=
		    &-\frac{1}{4}\left[X \partial_{X} \GG_{5}\right]\delta^{a c_{1}c_{2}c_{3}}_{b d_{1}d_{2}d_{3}}\xi_{c_{1}}\xi^{d_{1}} R_{c_{2}c_{3}}{}^{d_{2}d_{3}}\nonumber\\
		    &+\left[X\partial_{X}\GG_{3}+\partial_{\Phi}\GG_{4}+2X\partial^{2}_{X \Phi}\GG_{4}\right]\delta^{a c}_{b d}\xi_{c}\xi^{d}\nonumber\\
		    &+\frac{1}{2}\left[4\partial^{2}_{X\Phi}\GG_{4}+\partial_{X}\GG_{3}-\partial^{2}_{\Phi}\GG_{5}\right]\delta^{a c_{1}c_{2}}_{b d_{1}d_{2}}\xi_{c_{1}}\xi^{d_{1}}\nabla_{c_{2}}\Phi\nabla^{d_{2}}\Phi\nonumber\\
		    &+\left[\partial_{X}\GG_{4}+2X \partial^{2}_{X}\GG_{4}-\partial_{\Phi}\GG_{5}-X\partial^{2}_{X \Phi}\GG_{5}\right]\delta^{a c_{1}c_{2}}_{b d_{1} d_{2}}\xi_{c_{1}}\xi^{d_{1}}\Phi\nabla_{c_{2}}\nabla^{d_{2}}\Phi\nonumber\\
		    &+\left[\partial^{2}_{X}\GG_{4}-\partial^{2}_{X\Phi}\GG_{5}\right]\delta^{a c_{1}c_{2}c_{3}}_{b d_{1}d_{2}d_{3}}\xi_{c_{1}}\xi^{d_{1}}\nabla_{c_{2}}\nabla^{d_{2}}\Phi\nabla_{c_{3}}\Phi\nabla^{d_{3}}\Phi\nonumber\\
		    &-\frac{1}{2}[\partial_{X}\GG_{5}+X\partial^{2}_{X}\GG_{5}]\delta^{a c_{1}c_{2} c_{3}}_{b d_{1} d_{2}d_{3}}\xi_{c_{1}}\xi^{d_{1}}\nabla_{c_{2}}\nabla^{d_{2}}\Phi\nabla_{c_{3}}\nabla^{d_{3}}\Phi\\
		    \label{eq:dPPhiPhi_general_horndeski}
		    \delta \tilde{P}_{\Phi\Phi}(\xi)= &-[1+\partial_{X}\GG_{2}+2X \partial^{2}_{X}\GG_{2}+2\partial_{\Phi}\GG_{3}+2X\partial^{2}_{X\Phi}\GG_{3}]\xi^{2}\nonumber\\
		    &-[\partial^{2}_{X}\GG_{2}+2\partial^{2}_{X \Phi}\GG_{3}+2\partial^{3}_{X\Phi\Phi}\GG_{4}]\delta^{c_{1}c_{2}}_{d_{1}d_{2}}\xi_{c_{1}}\xi^{d_{1}}\nabla_{c_{2}}\Phi\nabla^{d_{2}}\Phi\nonumber\\
		    &-2[\partial_{X}\GG_{3}+X\partial^{2}_{X}\GG_{3}+2X\partial^{3}_{XX\Phi}\GG_{4}+3\partial^{2}_{X\Phi}\GG_{4}]\delta^{c_{1}c_{2}}_{d_{1}d_{2}}\xi_{c_{1}}\xi^{d_{1}}\Phi\nabla_{c_{2}}\nabla^{d_{2}}\Phi\nonumber\\
		    &-[\partial^{2}_{X}\GG_{3}+4\partial^{3}_{XX\Phi}\GG_{4}-\partial^{3}_{X\Phi\Phi}\GG_{5}]\delta^{c_{1}c_{2}c_{3}}_{d_{1}d_{2}d_{3}}\xi_{c_{1}}\xi^{d_{1}}\nabla_{c_{2}}\nabla^{d_{2}}\Phi\nabla_{c_{3}}\Phi\nabla^{d_{3}}\Phi\nonumber\\
		    &-\frac{1}{2}[\partial_{X}\GG_{4}+2X \partial^{2}_{X}\GG_{4}-\partial_{\Phi}\GG_{5}-X\partial^{2}_{X\Phi}\GG_{5}]\delta^{c_{1}c_{2}c_{3}}_{d_{1}d_{2}d_{3}}\xi_{c_{1}}\xi^{d_{1}} R_{c_{2}c_{3}}{}^{d_{2}d_{3}}\nonumber\\
		    &-\frac{1}{2}[\partial^{2}_{X}\GG_{4}-\partial^{2}_{X\Phi}\GG_{5}]\delta^{c_{1}c_{2}c_{3}c_{4}}_{d_{1}d_{2}d_{3}d_{4}}\xi_{c_{1}}\xi^{d_{1}}\nabla_{c_{2}}\Phi\nabla^{d_{2}}\Phi R_{c_{3}c_{4}}{}^{d_{3}d_{4}}\nonumber\\
		    &-\left[3\partial^{2}_{X}\GG_{4}+2X\partial^{3}_{X}\GG_{4}-2\partial^{2}_{X\Phi}\GG_{5}-X\partial^{3}_{XX\Phi}\GG_{5}\right]\delta^{c_{1}c_{2}c_{3}}_{d_{1}d_{2}d_{3}}\xi_{c_{1}}\xi^{d_{1}}\nabla_{c_{2}}\nabla^{d_{2}}\Phi\nabla_{c_{3}}\nabla^{d_{3}}\Phi\nonumber\\
		    &-[\partial^{3}_{X}\GG_{4}-\partial^{3}_{XX\Phi}\GG_{5}]\delta^{c_{1}c_{2}c_{3}c_{4}}_{d_{1}d_{2}d_{3}d_{4}}\xi_{c_{1}}\xi^{d_{1}}\nabla_{c_{2}}\nabla^{d_{2}}\Phi\nabla_{c_{3}}\nabla^{d_{3}}\Phi\nabla_{c_{4}}\Phi\nabla^{d_{4}}\Phi\nonumber\\
		    &+\frac{1}{3}[2\partial^{2}_{X}\GG_{5}+X\partial^{3}_{X}\GG_{5}]\delta^{c_{1}c_{2}c_{3}c_{4}}_{d_{1}d_{2}d_{3}d_{4}}\xi_{c_{1}}\xi^{d_{1}}\Phi\nabla_{c_{2}}\nabla^{d_{2}}\Phi\nabla_{c_{3}}\nabla^{d_{3}}\Phi\nabla_{c_{4}}\nabla^{d_{4}}\Phi\nonumber\\
		    &+\frac{1}{2}[\partial_{X}\GG_{5}+X\partial^{2}_{X}\GG_{5}]\delta^{c_{1}c_{2}c_{3}c_{4}}_{d_{1}d_{2}d_{3}d_{4}}\xi_{c_{1}}\xi^{d_{1}}\nabla_{c_{2}}\nabla^{d_{2}}\Phi R_{c_{3}c_{4}}{}^{d_{3}d_{4}}.
		\end{align}
	% section principal_symbol_for_the_most_general_horndeski_theory (end)
	\bibliographystyle{JHEP}
	\bibliography{hyperbolicity_general_horndeski}
\end{document}